\documentclass[runningheads]{llncs}
\usepackage{graphicx}
\usepackage{tabularx}
\usepackage{makecell}
\usepackage{xcolor}
\usepackage{xspace}

\newcommand{\project}{{GAIA Project}\xspace}
\newcommand{\labkit}{{\textit{LabKit}}\xspace}

\newcommand{\gaianode}{{\textit{GaiaNode}}\xspace}

\begin{document}


\title{Experiences from Using Gamification and IoT-based Educational Tools in High Schools towards Energy Savings\thanks{This work has been partially supported by the EU research project ‘‘Green Awareness In Action’’, funded by the European Commission and the EASME under H2020 and contract number 696029.This document reflects only the authors’ view and the EC and EASME are not responsible for any use that may be made of the information it contains.
\textbf{Preprint version of the paper submitted to 2019 European Conference on Ambient Intelligence, 13-15 November 2019, Rome, Italy}
}}

\titlerunning{Using Gamification and IoT-based Educational Tools in High Schools}
%
\author{Federica Paganelli\inst{1,2} \and
Georgios Mylonas \inst{3} \and Giovanni Cuffaro\inst{1} \and
Ilaria Nesi \inst{4}}
\authorrunning{ }
%
\institute{CNIT, Firenze, Italy  \\
\email{giovanni.cuffaro@cnit.it}\\
\and Computer Science Department, University of Pisa, Pisa, Italy\\
\email{federica.paganelli@unipi.it}\\
\and Computer Technology Institute and Press Diophantus, Greece \\ 
\and Gramsci Keynes High school, Prato Italy \\ \email{ilanesi@alice.it}
}
\maketitle              
\begin{abstract}
Raising awareness among young people, and especially students, on the relevance of behavior change for achieving energy savings is increasingly being considered as a key enabler towards long-term and cost-effective energy efficiency policies. However, the way to successfully apply educational interventions focused on such targets inside schools is still an open question. In this paper, we present our approach for enabling IoT-based energy savings and sustainability awareness lectures and promoting data-driven energy-saving behaviors focused on a high school audience. We present our experiences toward the successful application of sets of educational tools and software over a real-world Internet of Things (IoT) deployment. We discuss the use of gamification and competition as a very effective end-user engagement mechanism for school audiences. We also present the design of an IoT-based hands-on lab activity, integrated within a high school computer science curricula utilizing IoT devices and data produced inside the school building, along with the Node-RED platform. We describe the tools used, the organization of the educational activities and related goals. We report on the experience carried out in both directions in a high school in Italy and conclude by discussing the results in terms of achieved energy savings within an observation period.



\end{abstract}
\section{Introduction}

The Internet of Things (IoT) and smart cities have been two very active research fields during recent years, with a considerable amount of resources invested into building related infrastructures, creating large-scale smart city and IoT installations around the world. However, the question remains: how can we utilize such smart city and IoT deployments, in order to produce reliable, economically sustainable and socially fair solutions to create public value? This is especially true in the case of the educational domain, where it is more complex to integrate such solutions, given the restrictions in time and resources in school environments available for carrying out novel activities, in addition to traditional curricula and planned class schedules.

At the same time, there is an increasing interest in getting schools involved in raising awareness about climate change and energy efficiency. In this context, the importance of the educational community is evident, both in terms of size and future significance. Today's students are the citizens of tomorrow, and they should have the scientific and technological skills to respond to challenges like the climate change.  The relevance of sustainable energy and energy saving behavior is gaining increasing interest in schools and in educational programs. Indeed, in the last decade schools have been the target of studies, education initiatives as well as energy efficiency actions in several countries. There has also been interest recently regarding the design of educational activities for energy awareness centered around IoT-enabled experimentation approaches. 
It is a means through which Europe can meet its goals, by equipping citizens, enterprise and industry in Europe with the skills and competences  needed to provide sustainable and competitive solutions to the arising challenges~\cite{science-europe}.

In terms of research questions, given the context mentioned above, which we were interested to answer through our work, the first one would be ``how to engage and motivate end-user groups of students to participate in energy-saving educational activities''. The second one would be whether IoT-based and data-driven educational interventions towards sustainability awareness actually work inside the classroom. In other words, we wanted to look into the issue of motivating a school community using either more ``soft'' methods like gamification, or more technical hands-on ones, like IoT-based educational lab activities to educate students on the subject of sustainability.

Having in mind these questions, we discuss here our findings from applying gamification and competition mechanics in an Italian high school, with the purpose of getting the students more engaged into energy-saving activities in the context of a research project (from now on referred to as the \textit{Project}). We also present an educational toolkit, comprising IoT devices, data directly produced from school buildings, and the Node-RED programming environment. This toolkit allowed us to define educational activities within the school's computer science curriculum, leveraging IoT devices for environmental data acquisition and the integration with sensor measurements. We describe the organization of the activity and related curricula and educational goals to raise students' awareness in behaviour-based sustainability. Finally, we report on the experience carried out in both directions in a high school in Italy and conclude by discussing the results in terms of achieved energy savings in the observation period.

The activities described here were conducted in this school during 2 school years, 2017-18 and 2018-19. As mentioned above, they include the application of both ``soft'' (gamification and competition) and ``hands-on'' (lab kit) mechanisms in order to engage high school students. Section 4 discusses aspects related to gamification and competition, mostly related to activities in the schools during school year 2017-18, while Section 5 discusses aspects related to hands-on lab activities conducted during school year 2018-19. Our results indicate that both approaches can lead to interesting results; in our case, the increased engagement of the students led to both actual energy savings and positive learning outcomes.


\section{Related work}
\label{sec:related} 

The European Union is placing a strong focus on energy efficiency with initiatives like Build Up~\cite{build-up}, a portal for energy efficiency in buildings. Overall, the percentage of school buildings among non-residential ones in Europe is around 17\%~\cite{bpie11}. Regarding the current state of the art in inclusion of sustainability and other related aspects in the educational domain, there is a lot of activity taking place with respect to inclusion of makerspace elements in school curricula, aided by the availability of IoT hardware as well. \cite{PAPAVLASOPOULOU201757} summarizes recent activity within the Maker Movement approach, presenting relevant recent findings and open issues in related research. \cite{ERIKSSON20189} discusses a study stemming from a large-scale national testbed in Sweden in schools related to the maker movement, along with the inclusion of maker elements into the school curriculum of Sweden. 

Furthermore, there is a growing number of research projects and activities that focus specifically on energy efficiency within the educational domain such as ZEMedS~\cite{zemeds} and School of the Future~\cite{school-future}. 
Other recent projects like Charged~\cite{charged} and Entropy~\cite{entropy} target diverse end-user communities and do not focus specifically on the educational community. 
Moreover, several recent works focus on university curricula for teaching Internet of Things (IoT) leveraging a learning-by-doing and hands-on approach \cite{uni:zhamanov,uni:dobrilovic},\cite{uni:He},\cite{uni:abbasy}, while the design of IoT-enabled educational scenarios in high and junior schools is less investigated. Porter et al. \cite{school:porter} argue that the lack of students' engineering experiences in  primary and secondary education is in part due to the fact that very few teachers have an engineering/technology background and that the collaboration with universities and professionals would help coping with this issue. 
Along this direction, Gianni et al. \cite{gianni2018rapid} report on the usage of a toolkit \cite{mora2017tiles} for rapid IoT application prototyping with a group of high school students. Analogously, \cite{Katterfeldt201872} proposed the application in schools of a plug-and-play toolkit together with some suggestions for successful implementation of activities in high schools. An educational framework leveraging ubiquitous, mobile and Internet of Things technology for science learning in high schools has been proposed within the UMI-Sci-Ed project, while also investigating students' stance on IoT-enabled education activities \cite{umiscied}.
 
There are a few examples of IoT-driven educational activities performed with the additional objective of increasing students' awareness of societal challenges. Tziortzioti et al. \cite{tziortzioti2018raising} designed and experimented data-driven educational scenarios for secondary schools to raise students' awareness of water pollution. Mylonas et al. \cite{MYLONAS201943} proposed an educational lab kit and a set of educational scenarios primarily targeting primary schools for increasing energy awareness within the GAIA Project. \cite{Eriksson-2019-fablearn} is another recent example of a work discussing such issues from a more theoretic standpoint. 

With respect to gamification utilizing IoT in the context of sustainability, and specifically for energy and water, \cite{gamification-survey} is a recent survey on the subject. However, although there are several examples of using gamification in this context, there has been little focus so far on the benefits and dynamics of such an approach inside classrooms, an aspect we discuss in this work.
 


\section{Overview of the Deployment Environment}
\label{sec:general}


The work presented here was conducted in the context of a research project focusing on energy  efficiency in  educational  buildings, employing behavioural change strategies, i.e., not using invasive techniques or retrofitting the buildings with actuators. This project produced a real-world multi-site IoT infrastructure comprising several school buildings in Europe. 
The schools cover a  range  of  local  climatic conditions and educational levels (i.e., from primary to high school), as well as cultural settings (i.e., they are located in different countries). 

Within this infrastructure, hereafter referred to as \project IoT Platform, a large number of IoT monitoring endpoints have been installed installed inside classrooms using heterogeneous hardware  and  software  technologies, including different commercial hardware/sensor vendors, as well as open-source solutions. At each site, the following types of measurements are periodically acquired: the power consumption of the whole building and selected rooms/areas, the environmental parameters of selected classrooms and/or laboratories (typically 5-minutes average values), and weather  conditions  and  air  pollution  levels. These measurements, aggregated at different time granularity (e.g., 5-minute, hour, day, etc.), can be programmatically accessed through a set of REST APIs.

A set of software and hardware artifacts have been produced within the project with the aim of experimenting different ways for raising students awareness on sustainability and energy consumption topics, also leveraging data and services provided by the IoT platform made available by the project,
In this work we will focus on the following two software tools and the activities that have been designed around them:
\begin{itemize}
    \item a web application (from now on called the Challenge) that serves as a playful introduction to sustainability and energy-related concepts for students. It uses gamification mechanisms to increase end-user engagement;
    \item an IoT-based educational lab kit that has been setup using open-source technologies. The kit includes already assembled devices and commercial IoT sensors and actuators to allow students complete classes and lab tutorials regarding energy and sustainability.
\end{itemize}

In this work, we chose to focus on a specific high school located in Italy, where both types of activities have been performed. As mentioned in the introduction, the experiences relayed in this work cover a period of 2 school years. 

\section{Gamification and Competition}
\label{sec:challenge}

In this section, we report on our experience in supporting the use of the Challenge in three classes of the target school. Overall, the Challenge is an online application aimed at students, designed to raise energy awareness and act as a playful introduction to sustainability aspects by leveraging gamification mechanics. The core of the application is a set of online Quests, grouped into five subject areas related to energy consumption reduction. The Quests are offered to students as steps of a ``journey'', on top of a game ``board''. This journey can also be enhanced with Class Activities, which are designed by teachers. For instance, Class Activities can consist in deeper investigation of some topics, energy-saving actions in the real-life, observation of monitoring data provided by the IoT platform, etc. The educational activities supported by the Challenge aim at motivating participants to engage in energy saving topics, by seeing their impact on the school facilities’ energy consumption over the course of the challenge and, finally, competing against other classes and schools. 

\begin{figure}[h!]
\centerline{\includegraphics[width=0.98\columnwidth]{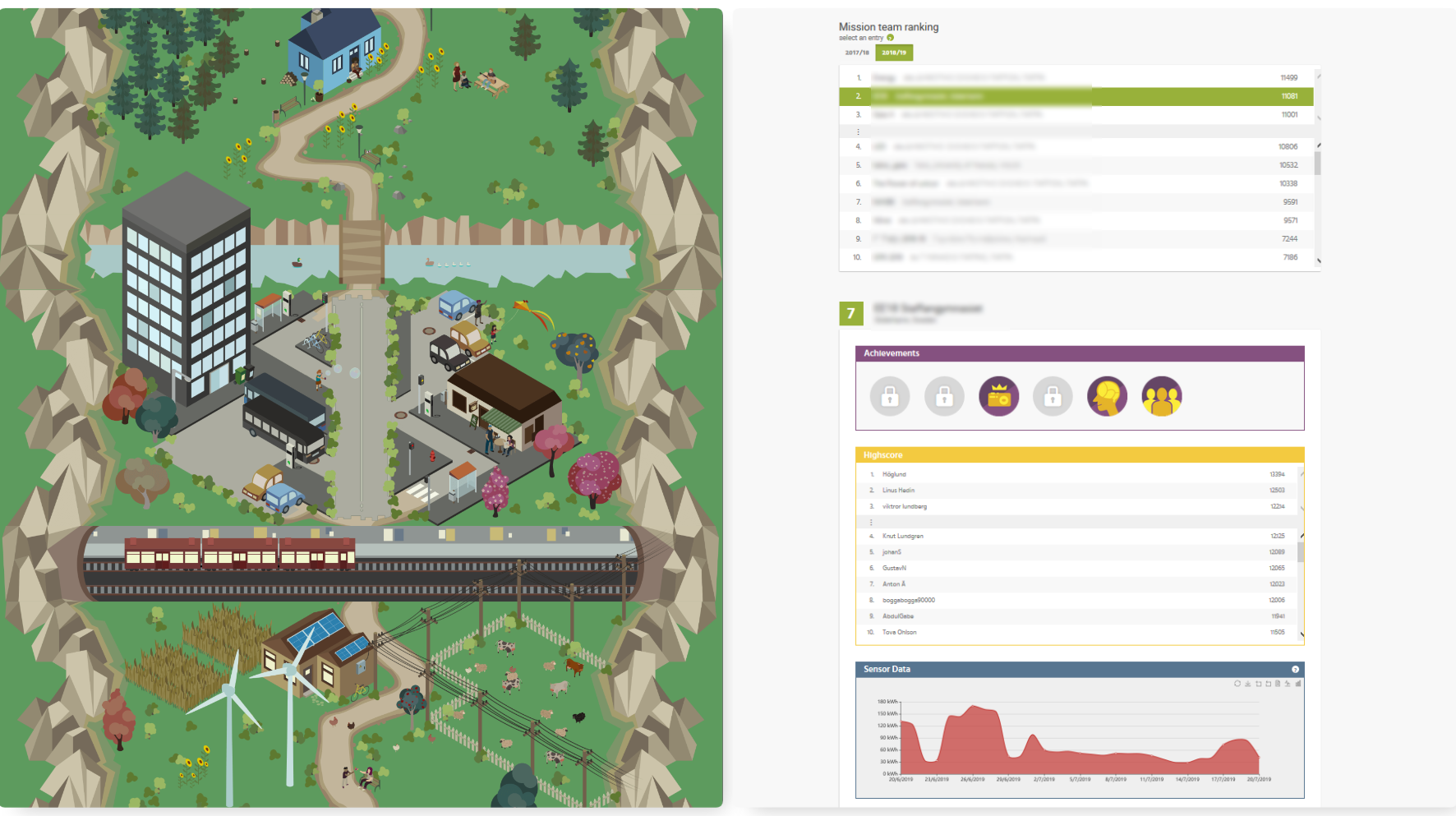}}
\caption{Some sample screenshots from the Challenge, with the game ``world'' on the left and the part where students see the schools' scores, trophies won and power consumption data from their school building on the right.}
\label{fig:challenge}
\end{figure}

Before introducing the Challenge to the classes, we performed a set of preparatory activities, such as disseminating advertising material in school areas and classrooms, training teachers through workshops, engaging the school principal and the technical staff. We also prepared brief eye-catching material, as an example of class activity created specifically for that school. Then, three classes of the high school started to participate to the activities. They conducted the Challenge Quests individually, while Class Activities were performed in groups. The students played the Challenge after a short introduction about the topics. Class Activities carried out by the classes consisted in analyzing monitoring data regarding the energy consumption and environmental comfort in the school facility, spotting possible rooms for improvement, devising and realizing ways of raising awareness in their school and family communities (e.g., news in the local newspaper, posts on the school web site, production of videos and presentations).


The students were divided into 5 groups and, to increase their engagement, students also underwent an evaluation. The teacher decided to assign a mark for the activity that is then accounted in the computation of the student's final mark for the subject (i.e., physics in this specific example). The way this was computed was interesting because it is based on a peer evaluation mechanism, and it was also proposed by the teachers of the school: first, a grade is assigned to the group as the average of the grades given by the other groups; then a grade is assigned for each member of a group by the other members of the same group. Student's grade is the weighted average of these two evaluations (25\% personal grade and 75\% group grade). Such an evaluation mechanism boosted the individual contribution to the group activity and fair cooperation among group members.

All students of the three classes took part to the Challenge's competition and the three classes started to climb on the Challenge's ranking of participating classes from all the schools participating in the Challenge (over 20 schools in total). They also produced some snapshots like animated GIFs and used a lot the portfolio function of the web application. Overall, the level of engagement due to competition achieved in the school was very high. This was a successful outcome, which, however, can be associated to some unexpected and risky effects, as we experienced in the final days of the competition. 

Indeed, the classes began to \textit{continuously monitor} the ranking and at a certain point in time they began arguing on fairness in the Challenge score and behavior of competitors. This issue was raised when they noticed that after months of activities, another surpassed them in the overall score. Essentially, what the students then tried to do amounted to ``reverse engineering'' the way that the scores were calculated in the Challenge. They basically started to monitor what the other teams were doing, and whether actions from their side had any tangible effect on their school's overall score. At some point, they limited the possible cases and scenarios for the way scores are calculated. This turn of events could be summarized as an ideal for the Challenge: our end-users were more than just engaged, they were \textit{thrilled} to participate and out-compete other schools, which they hadn't even heard of before.

Another aspect was that they assumed the way the other school surpassed them in the score was a ``trick'', suggesting to us that the score counting method was ``unethical''. They noticed that 3 new users were added to a competing school, thus contributing extra points to the overall class score.  They were very sensitive to this issue and stated that they were also ready to cancel all the project-related activities, since they perceived the whole project as anti-pedagogical, or even unethical. From their point of view, they were not complaining because they were no longer the best team, but against what they perceived as not fair play. The students also proposed a solution to the score ``issue'', as perceived by them. 

Essentially, the problem was that there was an assumption from the students' side that new students could be added to a certain school after an initial period and that all students from the other schools had registered early on. As a way to counter such complaints in the future, we setup some measures to reduce the probability of cheating behaviours, namely: i) registrations are allowed only for a limited period, teachers may request an extension providing a motivation, ii) the teacher's guide to the Challenge has been enhanced asking the teacher to control the identity of users registered to their class.







\begin{figure}[h!]
\centerline{\includegraphics[width=\columnwidth]{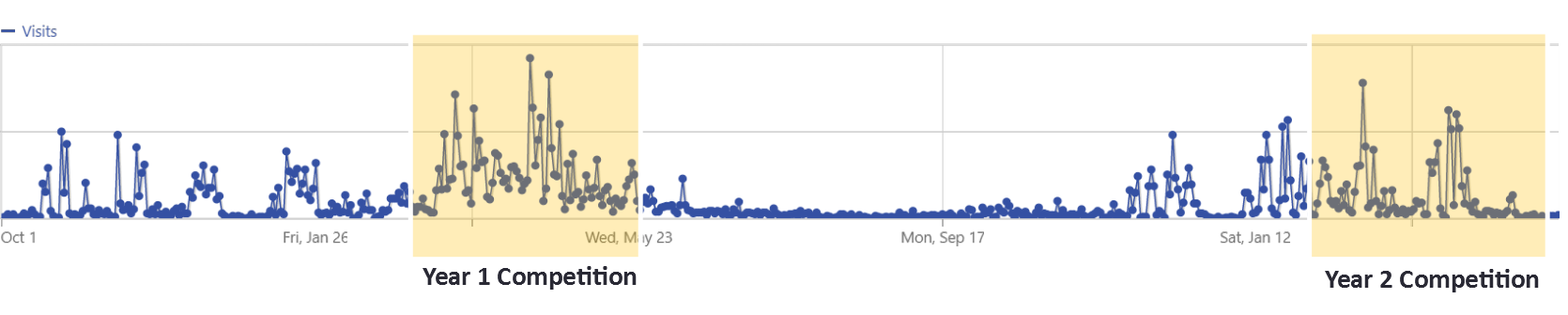}}
\caption{A chart depiction of the end-user visits to the Challenge, with the time periods of 2 competitions marked and showcasing the effect on engagement.}
\label{fig:challenge-stats}
\end{figure}

In terms of overall comments about the contribution of gamification to the engagement, apart from the aforementioned aspects in this specific school, in several of the other schools participating in the project we saw increased engagement. Moreover, during 2 school years (2017-18, 2018-19) we also announced 2 ``competitions'', where all schools in the project were called to participate and for certain categories they should use the Challenge to score more points or create content to share with their peers. As a general observation, during the time period that the competitions ran, there were very easily identifiable spikes in the end-user activity in the Challenge, as seen in Fig.~\ref{fig:challenge-stats}.

\section{IoT-enabled educational activity}
\label{sec:nodered}

In this section, we will focus on the experience carried out  in the design and experimentation of an "hands-on" IoT educational activity with the twofold objective of \textit{(i)} increasing students' awareness on the relevance of behavior change for achieving energy savings through a data-driven approach, thanks to real-world data gathered by the IoT platform made available by the \project, while also  \textit{(ii)} integrating such IoT-enabled experiment into a upper secondary school curricula. The second aspect, i.e. designing the activity so that it could fit within the educational program of at least of one the subjects offered to students was a key element to encourage teachers in devoting effort to contribute to the design and experimentation  as well as in actively engaging students in proficiently taking part to the activity.

The activity has been mainly designed to fit in the computer science curricula, although it can be further developed with actions carried out in the framework of additional subjects, such as sciences and physics. More specifically, the activity has been designed as a set of computer science lectures for a class of students of the 4th year of a Scientific Lyceum in Italy. The learning objectives for that year are defined at high level by the Minister of Education\footnote{https://www.miur.gov.it/liceo-scientifico-opzione-scienze-applicate (in Italian)} and then customized by each teacher. They generally consists in: learning a programming language, basics of coding, data modeling and tool for data access, manipulation and persistence, web programming. Moreover, the use of the acquired knowledge to support a study on topics in science and physics subjects is also welcomed.

In this context, the activity was designed to develop the theme of sustainability and energy awareness by first providing students with monitoring data gathered from the environment they live in (e.g. the school hall, the computer science laboratory,etc.) by using the sensor infrastructure deployed within the \project in a selected set of rooms and areas in the school (see Section \ref{sec:general} and a sensor board (called \labkit) that was given to students to gather environmental parameters in some of the remaining areas. Thanks to these tools, first, the students have access to a real-world dataset, which they can learn to manipulate and query through appropriate tools in order to derive meaningful data about the school environment they live in, related to comfort, resource usage, etc.. Based on such findings, students, with the help of the teachers, can decide on actions for further investigation (e.g., changing their behaviour to save energy, studying the factors that influence the comfort in school environments, etc.). Second, they can enhance the experience with programming tools to develop programs for data manipulation, sharing and visualization.

The material we prepared for computer science teachers consists in: the \labkit sensor board, documentation and examples for using a programming environment for developing programs for pushing sensor data from the \labkit to the IoT platform maintained in the \project and  simple web applications for data visualization. Hereafter, we describe the \labkit and the programming environment selected and extended for its usage within the \project and, then, report on the experience with a computer science class of 22 students.

\subsection{Lab Kit and programming tools}

The \labkit has been assembled utilizing the following components to minimize costs and ease replicability: 
\begin{enumerate}
    \item Raspberry Pi model 3B or 3B+: Raspberry Pi v3; 
    \item GrovePi, which is an add-on board that couples to the Raspberry to ease  the connection of external sensors;
    \item GrovePi Sensors, i.e., light, temperature, humidity, sound sensors, together with buzzer, LEDs and a Button; and 
    \item  an LCD screen, allowing an immediate feedback on gathered measurements.
\end{enumerate}

We selected Node-RED \cite{nodered} as a programming environment to be used by students since it is a well known environment that leveraged a flow-based programming environment for easily interconnecting hardware devices, online services and developing IoT applications.
Since Node-RED is a visual tool that allows users with minimal programming skills rapidly assembling and deploying an IoT application, it can be used in schools for performing simple experiences with sensors and IoT data processing in classes. Integrating Node-RED with the APIs of the \project IoT platform, and, optionally, local sensor kits, thus allows enriching the educational activities with the use of real data gathered from the sensor infrastructure, while leveraging a tool supported by a wide open source community and a rich documentation.

Node-RED comes with a set of ready-to-use customizable nodes allowing to design and deploy simple applications by simply dragging and dropping nodes from the palette, configuring and connecting them according to the desired flow. On the other side, Node-RED can also be easily extended by developing and adding new nodes to the repository.  This extensibility provides a flexible support to the design of educational activities at a customizable difficulty level, which can be adapted to the class level and syllabus. For instance, teachers may assign tasks to students for programming a custom data processing function or graphical widget and add them to Node-RED.
 In order to foster a data-driven approach, not merely limited to the local usage of the measurements gathered through the \labkit, an additional Node (called \gaianode) to access sensors and API of the \project IoT Platform as data sources in Node-RED. The software and documentation is available as open source on GitHub. 
The \gaianode plug-in for Node-RED is a set of nodes that allows interacting with the IoT platform of the \project to retrieve  measurements gathered by the fixed sensor infrastructure deployed in the schools involved in the \project, as well as pushing values of  the \labkit sensors  (e.g., Raspberry Pi sensors) into the platform.


\subsection{Data-driven Education for energy awareness}

Hereafter we describe one of the educational activities that have been proposed targeting high schools and integrating with the existing computer science curricula. The pedagogical goals aiming at increasing students awareness on energy topics are: awareness, observation, experimentation and action. This goals have been used to inform an IoT-enabled education activity, whose main steps are summarized in Table \ref{tab:activity}.

\begin{table} [ht!]
\caption{Template of IoT-enabled educational activity in high schools}
\label{tab:activity}
\begin{center}
\begin{tabular}{|m{0.06\textwidth}|p{0.47\textwidth}|p{0.47\textwidth}|} 
 \hline
 \textbf{Day} & \textbf{Activity description} & \textbf{Educational Goal}
\\\hline

1 & Introductory seminar organized by \project partners about the topics of energy consumption awareness and carbon footprint calculation & Introduction
\\\hline

2 & Introduction to the IoT. Definition and examples of deployment and  applications from the Web and the \project & Understanding the concept of IoT and  related impact in the everyday life
\\\hline

3 & Introduction to Node-RED and to the paradigm of Flow-based programming. Notion of node, flow and deployment in Node-RED  & Position the Flow-Based Programming paradigm using Node-RED for programming simple applications for data manipulation and delivery.
\\\hline

4 & Design of some basic Node-RED flow examples. Use of the \gaianode plugin for accessing IoT resources and measurements. Retrieving energy consumption and energy data of the school & Understanding relevance of extending Node-RED with new  nodes. Observation of measurements of energy consumption and environmental parameters in the school.
\\\hline

5 & Configuration of Raspberry and temperature sensor to monitor the temperature in a selected area (e.g., the computer science laboratory. & Learning setting up and configuring sensor and computing hardware devices. Analyzing and processing measurement data through a spreadsheet
\\\hline

6 & Develop a Node-RED flow application for creating a virtual sensor resource and pushing \labkit measurements into the \project platform  &  Learning programming a virtual sensor application pushing sensor measurements into the \project IoT platform.
\\\hline

7 & Develop a web-based Dashboard application to visualize temperature values & Developing programs for data access and visualization leveraging web protocols
\\\hline

8   & Analysis of monitoring data in different conditions of the lab (lights on/off, windows open/closed, etc. & Experimentation: taking some actions and analyzing the impact on the environment
\\\hline

9 & Discussion on findings and plan of short-term/long-term actions for further investigation, development, energy saving etc. & Analyzing data, finding issues and countermeasures, assigning priorities
\\\hline

10  & Action (objective and duration  of actions to be decided by students) & It depends on the action decided by the class
\\\hline
\end{tabular}
\end{center}
\end{table}

The educational activity has been designed by 
researchers with a computer networks background, with the help of the computer science professor of the high school 
where the activity was carried out. A total number of twenty two students of the high school participated in the activity. The activity has been carried out weekly in the 2-hour slot of computer science classes from February to end of April 2019. The students chose to monitor the temperature of their computer science laboratory, since they experienced a too high and uncomfortable heat. The availability of the \labkit allowed them monitoring the environmental conditions of the lab and correlating it with outdoor weather conditions retrieved through the \project IoT Platform. They measured very high temperature values (in the range of $25-30^\circ$C) also in cold days and during night, when heating was supposed to be off. They also analyzed these data while varying the room conditions (windows on/off, curtains open/closed).

Since radiators in the laboratory were not equipped with thermostatic valves, they couldn't turn their observations into direct energy saving actions (e.g., regulating radiators). As an outcome of the discussion on Day 9 they elicited a set of questions and energy-saving proposals and decided to submit them to the school principal. This resulted in a 20-minutes discussion with the principal on pragmatic actions for guaranteeing comfort while achieving energy savings. The discussion was initially focused on the experimental findings in the computer science laboratory and, at the end, was extended to other critical areas of the school. The discussion ended up with a set of actions to be performed by the school principal and ideas for follow-up activities to be performed by students.

\subsection{Evaluation}
\begin{figure}[h!]
\centerline{\includegraphics[width=0.98\columnwidth]{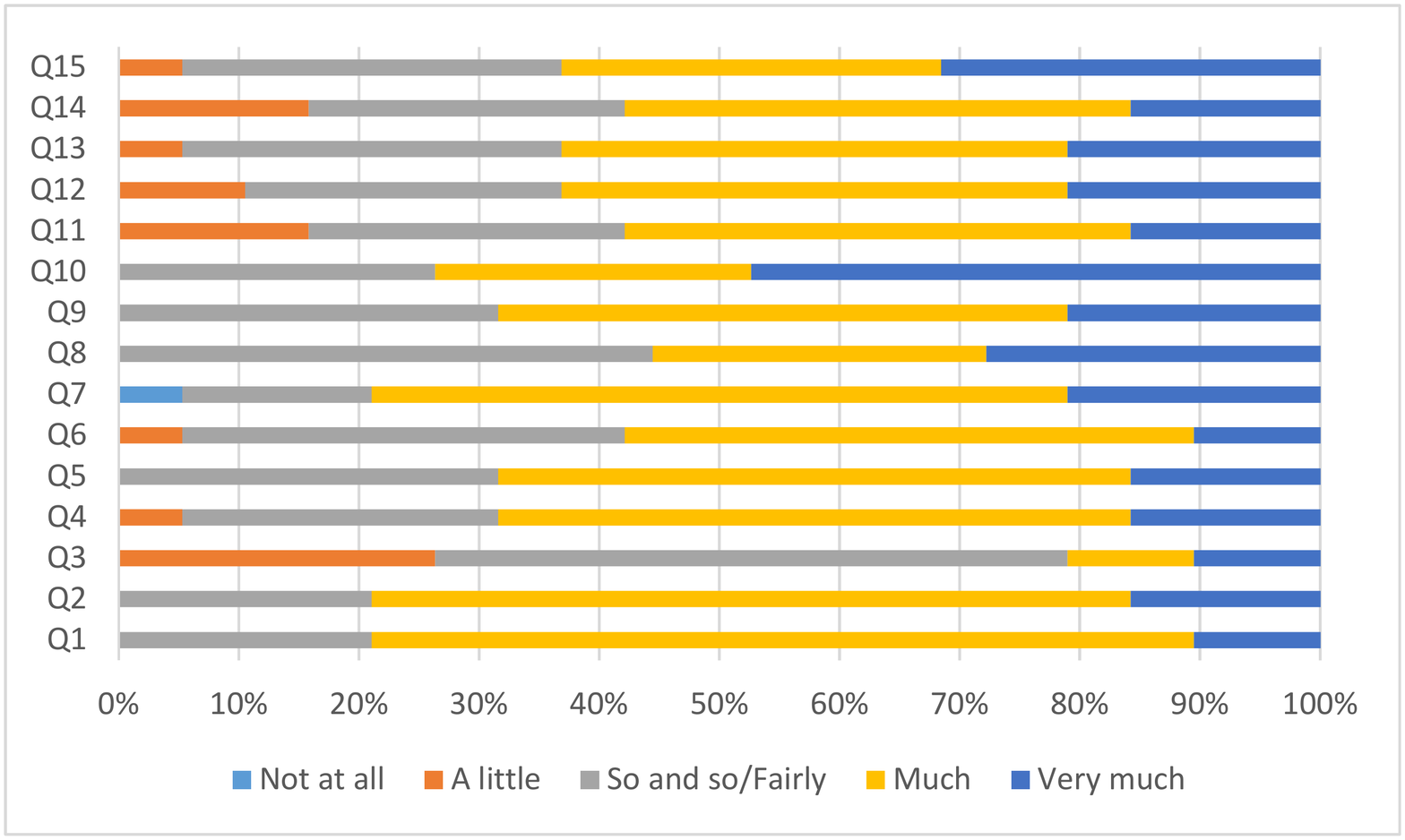}}
\caption{Results on the questionnaire related to the learning outcome and the experience}
\label{fig:quest_results}
\end{figure}

We gave a questionnaire to students to assess their satisfaction and engagement. The organization of the questionnaire is provided in Table \ref{tab:quest} and the statements were derived also taking into account analogous surveys in related work (e.g. \cite{umiscied} \cite{gianni2018rapid}. Answers were given in a Likert scale form (1-5). The questionnaire was submitted to the class after the end of the activity. Fig. \ref{fig:quest_results} shows the obtained results. The responses were mostly positive about the satisfaction and engagement in carrying out the activity (i.e., Q1,Q2 Q7,Q9 and Q10). However, responses related to the easiness of the activity (Q6 and Q8) suggest the need for improvements (e.g., distributing the activities within a longer time span, additional documentation/tutorials, etc.).


\begin{table} [h!]
\caption{Students' Questionnaire}
\label{tab:quest}
\begin{center}
\resizebox{1\columnwidth}{!}{
\begin{tabular}{|l|l|} 
 \hline
\textbf{ID} & \textbf{Question} \\
\hline
Q1 &	I am satisfied with the activity \\
\hline
Q2 &	I am pleased with the activity\\
\hline
Q3	& The activity was easy\\
\hline
Q4	& The process of the activity was clear and understandable\\
\hline
Q5 &	I was able to follow the tasks of the activity\\
\hline
Q6 &	I have the knowledge and ability to follow the tasks of the activity\\
\hline
Q7	& Attending the activity was enjoyable\\
\hline
Q8	& Attending the activity was exciting\\
\hline
Q9	& I was feeling good in the activity\\
\hline
Q10	& I found the activity useful\\
\hline
Q11	& The activity improved my capabilities in science and technologies\\
\hline
Q12	& I liked to observe and use the data and measurements\\
\hline
Q13	& I liked the lab activity with Node-RED and the Raspberry Pi\\
\hline
Q14	& I learned something new by observing and using the data and measurements\\
\hline
Q15	& I learned something new in the lab activity with Node-RED and the Raspberry Pi \\
\hline

\end{tabular}
}
\end{center}
\end{table}

Finally, since this activity was part of the computer science education program, at the end of the activity the students had to take an exam for assessment purposes. 
The exam was a written one and was made of 5 free-text questions (two questions on IoT and flow-based programming concepts, two on Node-RED usage for programming applications, and one question on solutions for sustainability and energy awareness in the school). The scores obtained by the class were are: 2 excellent, 3 good, 6 satisfactory, 7 sufficient, 4 insufficient. 
On average the class performed better with respect to previous computer science were better than previous exams. According to the teacher, this was probably due to the fact that previous exams have the aim of verifying knowledge acquired through traditional learning methodology on a larger body of content, while our IoT activity required more engagement and part of the content was produced by students themselves. In addition, the IoT activity had the peculiarity that the work was done in group and the collaboration was an incentive for students to perform better and to meet the deadline in order not to damage the other classmates.
As a final comment on the activity, the teacher also said that activity was successful in consolidating the relations among students and between class and the teacher. The students have demonstrated an increased engagement with respect to traditional lectures in previous months, thanks also to the awareness that their work was going to have impact on the school environment where they live in. As lesson learnt, the teacher also suggested that the timeplan was too strict and the activity would benefit of additional time resources (e.g., three hours a week instead of two) devoted to the activity. In that specific case, this would mean performing part of the activity within other subjects (physics or science), thus requiring the cooperation of a group of teachers and the enhancement of interdisciplinary aspects.

\section{Behaviour-based Energy Savings}
\label{sec:savings}



Since the common objective of the educational activities presented above is energy awareness, in this section we report on an activity planned and performed by students targeting energy savings in a school environment through behaviour change. The following is an example of possible decisions that students can take as an effect of awareness improvement through learning, observation and experimentation steps, as the ones described in Sections \ref{sec:challenge} and \ref{sec:nodered}.



\begin{figure}[ht]
\centerline{\includegraphics[width=0.9\columnwidth]{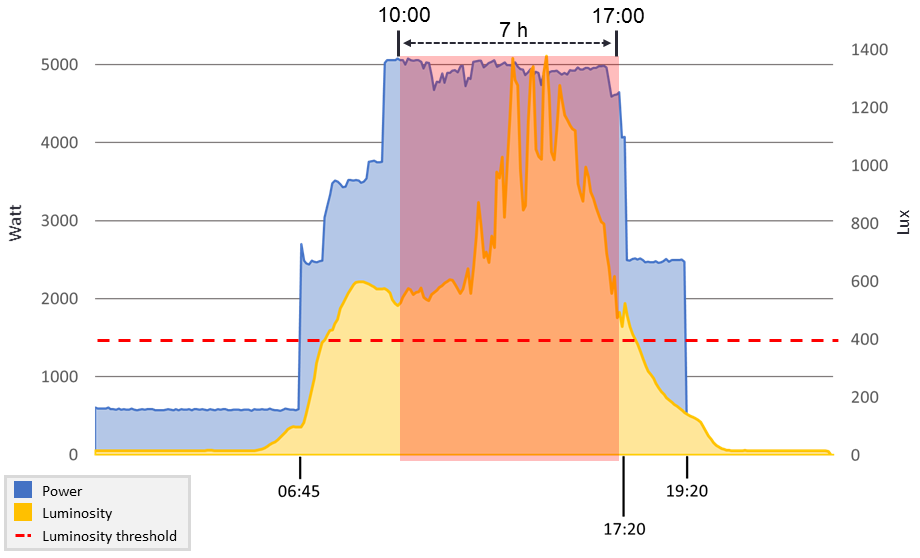}}
\caption{Graph with light level threshold and with the period in which there is a waste of electricity highlighted}
\label{fig:prato1}
\end{figure}

The class decided to focus on the lighting of the main hall as the use-case for targeting energy savings. With respect to luminosity, there is a minimum recommended value of 150 lux for a circulation area as an indoor hall. There are a number of luminosity sensors installed in this specific building. Given that the sensors produce that are highly related to their orientation, which is not optimal for calculating a luminosity average value, the students had to approximate the values they saw through the system.  Making a rough estimation, students set a threshold of 400 lux for the values produced by the sensors that they thought it corresponded to ``good enough'' lighting. Figure~\ref{fig:prato1} displays the measurements for power consumed by lights in the hall and luminosity, with the addition of the 400 lux threshold (horizontal line marked in red). Also highlighted in the figure is the interval during which luminosity in the school hall is above the threshold. 

It is evident that between 10:00AM and 5:00PM the lights should be turned off. This is a recurring situation in this specific school building for a number of months, due to its location (Mediterranean) and orientation; i.e., it is not something that is observed only for a single day or over a short time period. The next step was to act on the plant for turning off the unnecessary lights, while also making sure not to leave any part of the hall in the dark. Lighting should be turned off for sufficient time, in order to be able to observe the change in the data. It was convenient to calculate the average values of the lighting system during a ``normal'' baseline period and after the intervention.

The school analyzed the new data regarding power during the period in which the light was turned off. With the lighting configured as usual, power consumption is at approximately 4.9kW. When the school acted to keep active only what is necessary, the power consumption decreased to 1.9kW, thus saving 3kW in the process. This practically means that 21kWh could be saved during a single day, considering the 7 hours of the interval during which this issue was identified. With such data in hand, students performed simple actions for raising awareness in the school staff for switching off the lights in the hall when not necessary and involved their schoolmates in similar actions in classrooms and laboratories.

\begin{figure}[h!]
\centerline{\includegraphics[width=\columnwidth]{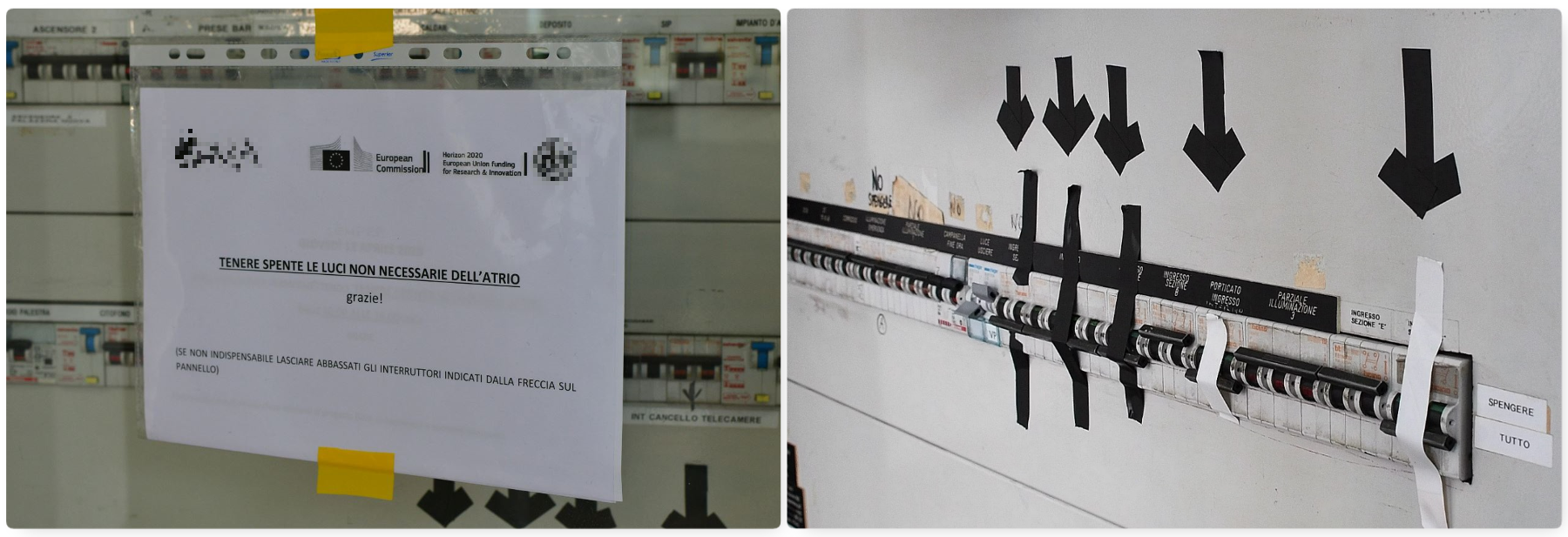}}
\caption{Some examples of the simple actions/interventions that the students made in the school building.}
\label{fig:student-actions}
\end{figure}

The potential energy savings analyzed in the previous steps pushed the students to act. They created a set of signs (Fig.~\ref{fig:student-actions}) aiming at helping the school staff to remember which switches can be turned off when the natural light is enough. They also designed a poster to be placed in all the rooms equipped with a projector by the students: ``Please shutdown the computer and the projector when not in use''. Additionally, one of the classes involved produced a short video to encourage their friends and families to join the ``battle for environmental care''. This short video gives some simple advice for saving energy and decreasing pollution.

Based on achievements due to the short- and medium-term experimental activities, the school wanted to support students in taking further measures within a longer time span, to obtain the best results in terms of energy efficiency. Students observed and analyzed periodically the impact that these changes have in the long term and monitored progress toward the achievement of their objectives.


\section{Conclusions and future work}

The educational community is one of the most interesting target groups for sustainability and energy savings-related activities. The successful introduction of such activities into the curriculum of schools in Europe is still an open issue. In this paper, we have presented our experiences from utilizing mechanisms such as competitions, gamification and IoT-based, hands-on lab activities to increase the engagement of students at a high school in Italy. 

Our results, produced during a long time period covering two school years, provide some interesting insights regarding the creative ways the aforementioned mechanisms can be used to trigger the interest of high school students. The inclusion of competition and gamification aspects when having students as end-users can increase their engagement rapidly, especially when having students groups or schools compete with each other. Some practical guidelines from our experience are the following:

\begin{itemize}
\item Direct and informal support to teachers: teachers were the gateway to our end-users; having gained the trust and attention of teachers is the first step to establishing a connection with the students as well.
\item Provide short and captivating material: schools tend to have little time available to dedicate to extra-curricular activities, information should be as engaging and codified as possible.
\item Competitiveness is key for engagement but has to be handled with care: in our case, a rapid increase in engagement came close to backfire, as discussed in Section 4.
\end{itemize}


With respect to our future work, we plan to continue this line of research by applying our tools to other school communities in other European countries.

\section{Acknowledgement}

 \bibliographystyle{splncs04}
 \bibliography{biblio}

\end{document}